\documentclass[12pt]{article}
\usepackage{pst-plot,epsf}
\newcommand{\AmS}{{\protect\the\textfont2
  A\kern-.1667em\lower.5ex\hbox{M}\kern-.125emS}}

\newcommand{\beq}{\begin{equation}}
\newcommand{\eeq}{\end{equation}}
\newcommand{\bea}{\begin{eqnarray}}
\newcommand{\eea}{\end{eqnarray}}

\newcommand{\epm}{e^+e^-}

\newcommand{\ra}{\rightarrow}

\newcommand{\eett}{e^+ e^- \ra t \bar{t}}
\newcommand{\ttbar}{t\bar{t}}
\newcommand{\bwbw}{b W^+ \bar{b} W^-}

\newcommand{\eesixf}{e^+ e^- \ra b f_1 \bar{f'_1} \bar{b} f_2 \bar{f'_2}}
\newcommand{\eebnmbdu}{e^+ e^- \ra b \nu_{\mu} \mu^+ \bar{b} d \bar{u}}
\newcommand{\eebudbdu}{e^+ e^- \ra b u \bar{d} \bar{b} d \bar{u}}
\newcommand{\eebnmbmn}{e^+ e^- \ra b \nu_{\mu} \mu^+ \bar{b} \mu^- 
                                                          \bar{\nu}_{\mu}}

\newcommand{\bnmbmn}{b \nu_{\mu} \mu^+ \bar{b} \mu^- \bar{\nu}_{\mu}}
\newcommand{\bnmbdu}{b \nu_{\mu} \mu^+ \bar{b} d \bar{u}}
\newcommand{\budbmn}{b u \bar{d} \bar{b} \mu^- \bar{\nu}_{\mu}}
\newcommand{\bnmbtn}{b \nu_{\mu} \mu^+ \bar{b} \tau^- \bar{\nu}_{\tau}}

\begin{document}
\thispagestyle{empty}
\begin{flushright}
October 2002\\
\vspace*{1.5cm}
\end{flushright}
\begin{center}
{\LARGE\bf Top quark pair production and decay at linear colliders: 
             signal vs. off resonance background\footnote{Talk 
           presented by K. Ko\l odziej at 6th International Symposium 
           on Radiative Corrections 
          ``Application of Quantum Field Theory to Phenomenology -- 
           RADCOR~2002'' and 6th Zeuthen Workshop on Elementary Particle 
           Theory ``Loops and Legs in Quantum Field Theory'', Kloster Banz, 
           Germany, September 8--13, 2002.}}\\
\vspace*{2cm}
Adam Biernacik\footnote{Work supported in part by the Polish State 
           Committee for Scientific Research (KBN) under contract No. 
           2~P03B~035~22.} and 
Karol Ko\l odziej\footnote{Work supported
           in part by the Polish State Committee for Scientific Research
           (KBN) under contract No. 2~P03B~045~23 and by European 
           Commission's 5-th Framework contract  HPRN-CT-2000-00149.}$^,$
           \footnote{E-mail: kolodzie@us.edu.pl}\\[1cm]
{\small\it
Institute of Physics, University of Silesia\\ 
ul. Uniwersytecka 4, PL-40007 Katowice, Poland}\\
\vspace*{3.5cm}
{\bf Abstract}\\
\end{center}
Standard Model predictions for the reactions with six fermions 
in the final state relevant for top quark pair production and decay at linear 
colliders are discussed. An issue 
of the double resonance signal versus non doubly resonant background is 
addresed.
Effects related to the off-mass-shell production of the $\ttbar$-pair are 
discussed.
\vfill
\newpage

\section{MOTIVATION}

As the top quark is the heaviest particle ever observed, the measurement 
of its physical properties  may give hints towards better 
understanding of the electroweak (EW) symmetry breaking mechanism and observed 
fermion mass hierarchy. Should effects of physics beyond the 
Standard Model (SM)
be visible at the energy scale below 1 TeV, then top couplings may show 
deviations from the corresponding SM values.
Therefore high-precision measurements of the top quark properties and 
interactions are planned at TESLA \cite{Tesla} and will most certainly belong 
to the research program of any future $\epm$ collider \cite{NLC}.

To disentangle the possible effects of physics beyond the SM, it is crucial 
to know the SM predictions for the top quark pair production and decay 
precisely, including radiative corrections.

The quantum chromodynamics (QCD) radiative corrections to the on-mass-shell 
top pair production
\beq
\label{eett}
         \epm \ra \bar{t} t
\eeq
are known both in the threshold region, to the next-to-next-to-leading 
order including the effects of initial state radiation and
beamstrahlung \cite{topNNLO}, and, in the continuum, to order $\alpha_s^2$
\cite{eettQCD}. The electroweak radiative corrections to reaction 
(\ref{eett}) are known to one-loop order \cite{eettEW}, \cite{BZ}. They are 
typically of $\mathcal{O}$(10\%).

The corrections to the top decay into a $W$-boson and a 
$b$-quark, and to the $W$ decay into a fermion pair are
known, too, to order $\mathcal{O}$$(\alpha\alpha^2_s)$ \cite{topdec2}
and to order $\mathcal{O}$$(\alpha\alpha_s)$ \cite{Wdec}, respectively.

As the $t$-quark ($\bar{t}$-quark) of reaction (\ref{eett}) almost immediately 
decays predominantly into a $b$-quark ($\bar{b}$-quark) 
and $W^+$ 
($W^-$), and the latter decays into a fermion pair, the 6 fermion 
reactions of the form
\bea
\label{eesixf}
         \eesixf,
\eea
where $f_1=\nu_{\mu}, \nu_{\tau}, u, c$, $f_2=\mu^-, \tau^-, d, s$ 
and $f'_1$, $f'_2$ are the corresponding weak isospin partners,
$f'_1=\mu^-, \tau^-, d, s$, $f'_2=\nu_{\mu}, \nu_{\tau}, u, c$,
should be studied, at least in the lowest order of SM, with 
a complete set of the Feynman diagrams \cite{KK}.

In the present lecture, the effects caused by off-shellness of the 
$\bar{t}t$ pair and by the off resonance background contributions
to reactions (\ref{eesixf}), which may become 
important, especially for measurements at the c.m.s. energies much 
above the threshold, will be discussed. We will present results on 
\bea
\label{eebnmbmn}
                 \eebnmbmn,
\eea
which is a pure EW reaction. In the unitary gauge, if the Higgs boson 
couplings to electrons and to muons are neglected, reaction (\ref{eebnmbmn})
receives contributions from 452 Feynman diagrams.
The results obtained with a full set of the diagrams will be 
compared with the double resonance approximation for 
a $t$- and $\bar{t}$-quark
\bea
         \epm \ra \bar{t}^* t^* \ra \bnmbmn, 
\eea
with only two signal diagrams contributing, and with the narrow width
approximation for the $t$- and $\bar{t}$-quark.

As it is not feasible to calculate 
radiative corrections to the full set of Feynman diagrams, it will 
be argued that it is reasonable to include higher order effects, at least the 
leading ones, just for the two signal diagrams.

\section{OUTLINE OF CALCULATION}

\subsection{A program {\tt eett6f}}

Matrix elements are calculated with the helicity amplitude method 
described in \cite{KZJ}.
Phase space integrations are performed with the Monte Carlo (MC) method.
The most relevant peaks of the matrix element squared, related to 
the Breit-Wigner shape of the $W, Z$, Higgs and top quark resonances,
and to the exchange of a massless photon or gluon,
have to be mapped away. 

As it is not possible to find out a single parametrization of the 
$14$-dimensional phase space which would allow to cover the whole 
resonance structure of the integrand, it is necessary to utilize 
a multichannel MC approach.

A computer program {\tt eett6f} for calculating cross sections of 
reactions (\ref{eesixf}) at c.m.s. energies 
typical for linear colliders has been written. Version 1.0
of {\tt eett6f} \cite{program} allows for calculating both total and 
differential cross sections at tree level of SM. 
The program can be used as the Monte Carlo generator of unweighted
events as well.

{\tt eett6f} is a package written in {\tt FORTRAN 90}.
It consists of 50 files including a makefile, all stored in 
one working directory. The user should 
specify the physical input parameters in module {\tt inprms.f} and 
select a number of options in the main program {\tt csee6f.f}. 
The options allow, among other, for calculation
of the cross sections while switching on and off different subsets of the 
Feynman diagrams.
It is also possible to calculate cross sections in two different
narrow width approximations, for $\ttbar$-quarks, or $W^{\pm}$-bosons.
The program allows for taking into account both the electroweak 
and QCD lowest order contributions.

Constant widths of unstable particles, the massive electroweak vector 
bosons, the Higgs boson and the top quark, are introduced through the complex 
mass parameters
\bea
\label{masses}
 M_V^2 &\!\!=\!\!& m_V^2-im_V\Gamma_V, \quad V=W, Z, \\
 M_H^2&\!\!=\!\!& m_H^2-im_H\Gamma_H,  
                         \quad M_t=m_t-i\Gamma_t/2, \nonumber
\eea
which replace masses in the corresponding propagators, both in 
the $s$- and $t$-channel Feynman diagrams,
\bea
\Delta_F^{\mu\nu}(q)\!&=&\!\frac{-g^{\mu\nu}+q^{\mu}q^{\nu}/ M_V^2}
                               {q^2- M_V^2},   \\
\Delta_F(q)\!&=&\!\frac{1}{q^2- M_H^2}, \qquad
S_F(q)=\frac{/\!\!\!q+ M_t}{q^2- M_t^2}. \nonumber 
\eea
Propagators of a photon and a gluon are taken in the Feynman gauge.

The EW mixing parameter may be defined either real or complex,
\bea
\sin^2\theta_W=1-\frac{m_W^2}{m_Z^2}, \quad {\rm or} \quad
\sin^2\theta_W=1-\frac{M_W^2}{M_Z^2}.
\eea
As light fermion masses are not neglected, cross sections
can be calculated without any kinematical cuts.

\subsection{Checks}

A number of checks of a program {\tt eett6f} have been performed.

Matrix elements of different `subprocesses' of (\ref{eesixf}): 
$\eett$, $\bwbw$, $b f_1 \bar{f'_1} \bar{b} W^-$, $b W^+ \bar{b} f_2 
\bar{f'_2}$ and $f_1 \bar{f'_1} f_2 \bar{f'_2} Z$ have been checked against 
{\tt MADGRAPH} \cite{MADGRAPH}.
Moreover, matrix elements of the `subprocesses' with the single on-mass-shell 
top quark have been calculated in an arbitrary linear gauge \cite{BCK}.
Finally, matrix elements of reaction (\ref{eebnmbmn}) have been programmed 
independently by each of the two authors.

The multichannel phase space generation routine has been 
checked by comparing normalization of different channels 
against each other and testing energy-momentum conservation 
and on-mass-shell relations. Moreover,
results obtained with the MC integration routine {\tt VEGAS} \cite{vegas}
have been reproduced with a more efficient own made MC integration 
routine {\tt carlos}.

Results for different channels of reaction (\ref{eesixf}) have been
checked against those published in the literature.
A comparison against {\tt LUSIFER} \cite{DR} is shown in Table~\ref{tab:didi}.
The discrepancy of a few standard deviations between the results can most 
probably be explained
by slightly different implementation of the top quark width.
The invariant mass and angular distributions of the 
$bu\bar{d}$-quark triple of $\epm \ra \budbmn$ at $\sqrt{s}=500$ GeV 
of \cite{DR} are nicely reproduced by {\tt eett6f} within accuracy  
of plots.
\begin{table}
\begin{center}
\caption{\small Lowest order total cross sections for 3 different channels 
         of reaction  (\ref{eesixf}) with cuts of \cite{DR} and zero 
         external fermion masses}
\label{tab:didi}
\vspace{0.5cm}
\begin{tabular}{lcc}
\hline
\rule{0mm}{7mm} & \multicolumn{2}{c}{$\sigma(\sqrt{s} = 500 
                                         \;{\rm GeV})$ (fb)}\\[2mm]
\cline{2-3}
\rule{0mm}{7mm} $\epm \ra $     & \tt eett6f & \tt LUSIFER \\[2.0mm]
\hline
\rule{0mm}{7mm} $\bnmbmn$ & 5.8065(33) & 5.8091(49) \\[1.5mm]
                $\bnmbtn$ & 5.8196(32) & 5.7998(36) \\[1.5mm]
                $\bnmbdu$ & 17.275(28) & 17.171(24) \\[1.5mm]
\hline
\end{tabular}
\end{center}
\end{table}

Lowest order SM total cross sections of $\eebnmbdu$ agree nicely
with those of \cite{YKK}. Moreover, a qualitative agreement with the results of
\cite{ABPG} have been found, see \cite{KK} for details. 

\section{NUMERICAL RESULTS}

Lowest order SM total cross sections of reaction (\ref{eebnmbmn}) at 3
different c.m.s. energies typical for a linear collider are collected
in Table~\ref{tab:total}. The physical input parameters which are used are the
same as those of \cite{KK}. The cross section $\sigma$ obtained with the 
full set of the Feynman diagrams is shown in column 1, the cross section 
$\sigma_{\bar{t}^*t^*}$ obtained with the two $\ttbar$ signal diagrams only,
in column 2, and the cross section in the narrow top width approximation, 
$\sigma_{\bar{t}t}$, in column 3. The results differ between each
other by a few per cent at $s^{1/2}=360$ GeV and 500 GeV, while
at $s^{1/2}=800$ GeV the differences become even larger, of more than 10\%.

{\begin{table}[hbt]
\begin{center}
\caption{\small Lowest order SM total cross sections of (\ref{eebnmbmn})}
\vspace{0.5cm}
\label{tab:total}
\begin{tabular}{cccc}
\hline
\rule{0mm}{7mm} $\sqrt{s}$ & $\sigma$ & $\sigma_{\bar{t}^*t^*}$ & 
                                   $\sigma_{\bar{t}t}$ \\[2mm]
\cline{2-4}
\rule{0mm}{7mm}  (GeV) & \multicolumn{3}{c}{(fb)} \\[2mm]
\hline
\rule{0mm}{7mm}   360 & 4.416(6) & 4.262(1) & 4.624(2) \\[1.5mm]
  500 & 6.705(6) & 6.354(2) & 6.400(7) \\[1.5mm]
  800 & 3.538(29)  & 3.058(2) & 2.973(4) \\[1.5mm]
\hline
\end{tabular}
\end{center}
\end{table}
Dependence of the total cross section of (\ref{eebnmbmn}) on a width of
the top quark, $\Gamma_t$, treated as a free parameter, is shown in 
Table~\ref{tab:totgt}.
It is amazing, that despite the contribution of 450 off resonance Feynman
diagrams, we see substantial dependence on the top width, $\sim 1/\Gamma_t^2$,
even at c.m.s. energies much above threshold, a bahaviour
typical for the resonance $\ttbar$-pair production which proceeds through
the two signal diagrams only.
This may be helpful in determining 
$\Gamma_t$ from measurements of the total top pair production cross section 
in the continuum, complementary to the measurement of the top quark width 
at the $\ttbar$ threshold.

\begin{table}[hbt]
\begin{center}
\caption{\small Lowest order SM total cross sections of (\ref{eebnmbmn}) in fb 
         for different values of the top quark width $\Gamma_t$}
\label{tab:totgt}
\vspace{0.5cm}
\begin{tabular}{cccc}
\hline
\rule{0mm}{7mm} $\sqrt{s}$  & \multicolumn{3}{ c}{$\Gamma_t$ (GeV)} \\[2mm]
\cline{2-4}
\rule{0mm}{7mm}  (GeV) & 1.5 & 1.6 & 1.7 \\[2mm]
\hline
\rule{0mm}{7mm}  360 & 4.416(6) & 3.862(7)  & 3.423(7) \\[1.5mm]
 500 & 6.705(6) & 5.942(9) & 5.296(8) \\[1.5mm]
 800 & 3.538(29)& 3.130(14) & 2.785(11)\\[1.5mm]
\hline
\end{tabular}
\end{center}
\end{table}

\rput(7,-6){\scalebox{0.8 0.8}{\epsfbox{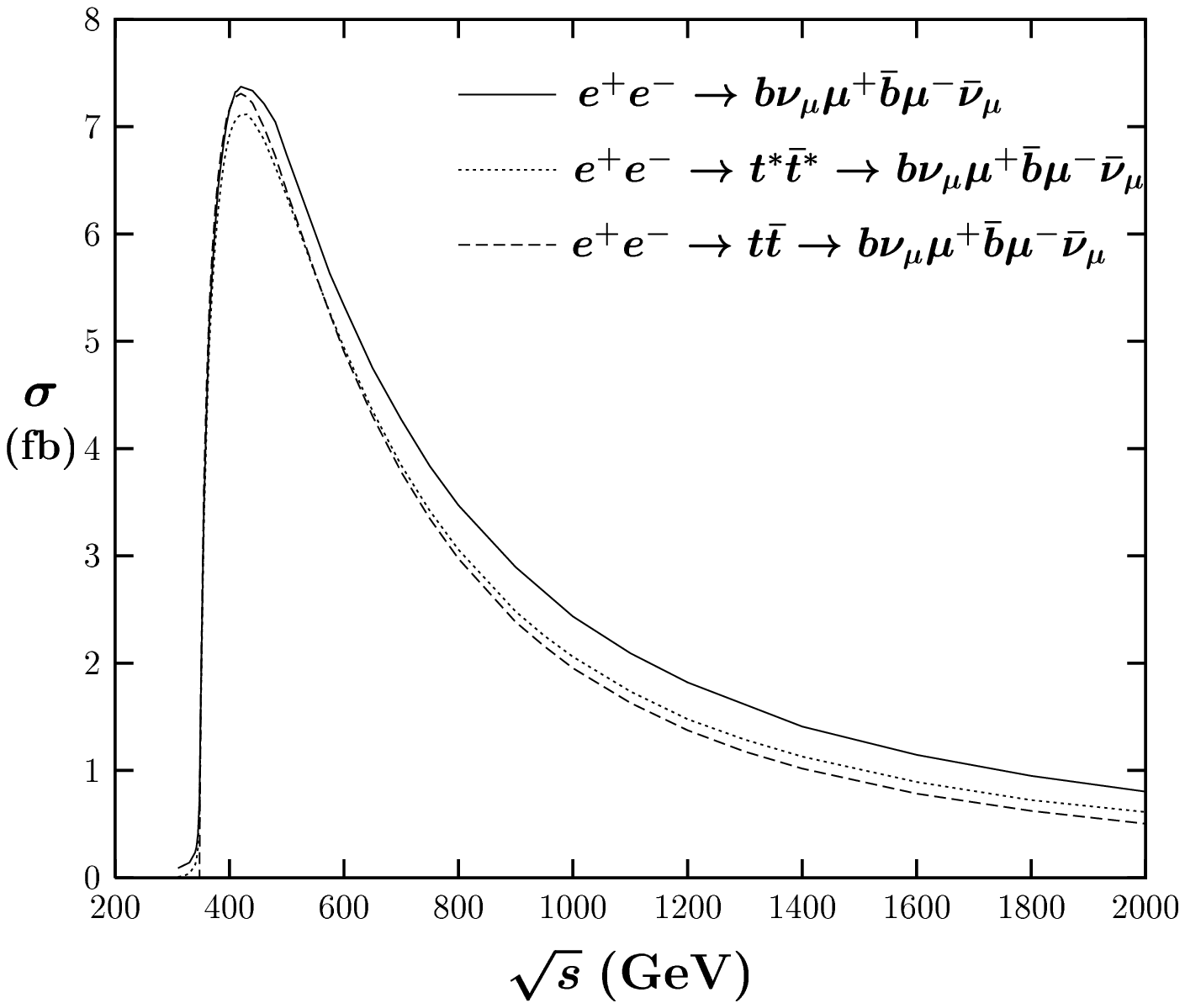}}}

\vspace*{9.0cm}

\bigskip
\bigskip
\begin{center}
{\small Figure~1. Total cross sections of $\eebnmbmn$ as functions of the 
                  c.m.s. energy.\hspace*{1.cm}}
\end{center}

\rput(7,-6){\scalebox{0.8 0.8}{\epsfbox{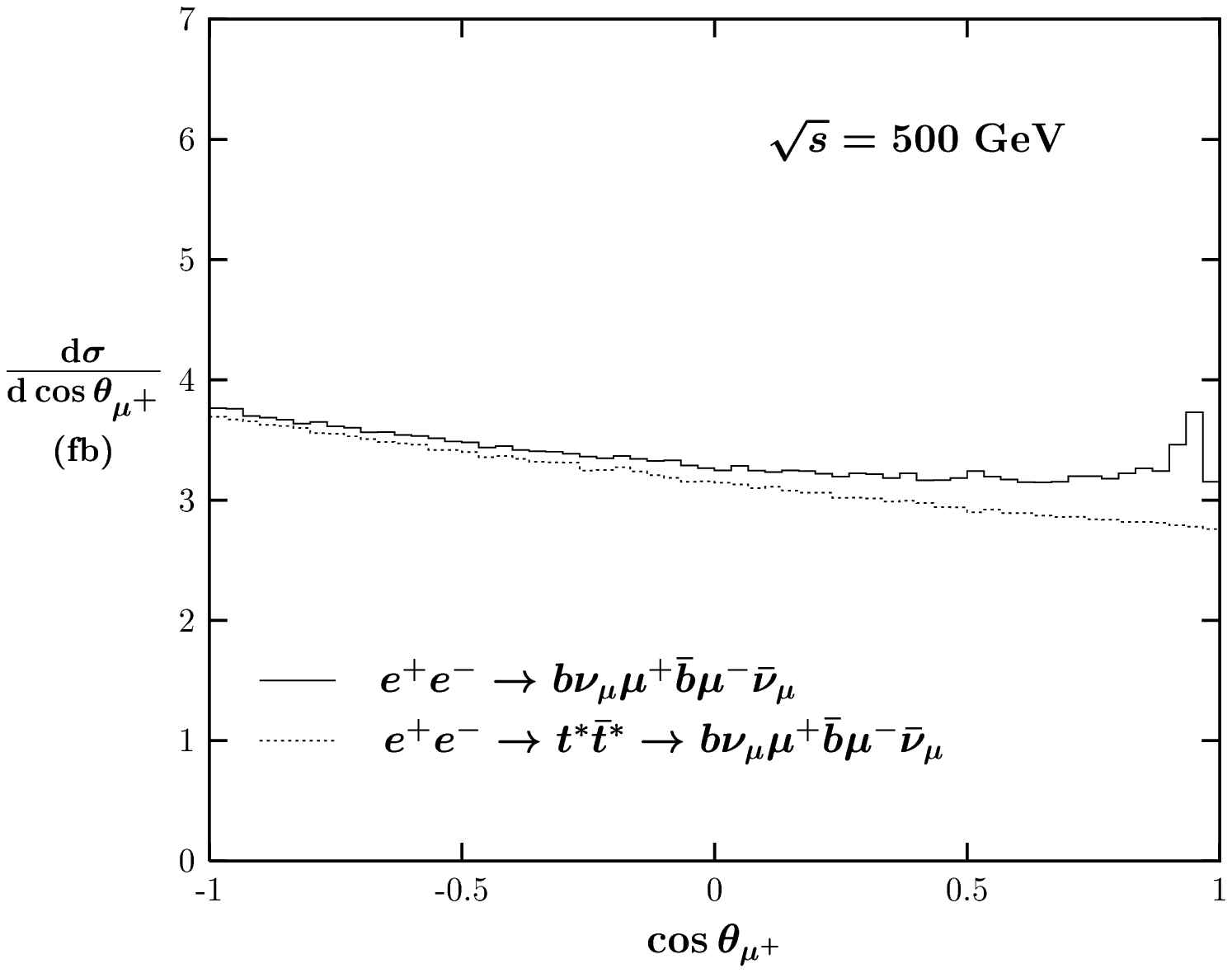}}}

\vspace*{9.0cm}

\bigskip
\bigskip
\begin{center}
{\small Figure~2. Angular distributions of a $\mu^+$ at $\sqrt{s}=500$ GeV.
\hspace*{2.cm}}
\end{center}

The energy dependence of the total cross sections of (\ref{eebnmbmn}) is 
shown in Figure~1. The full lowest order cross section $\sigma$ 
is plotted with the solid line, the signal cross section 
$\sigma_{\bar{t}^*t^*}$ with the dotted line, and the cross section in 
the narrow top width approximation, $\sigma_{\bar{t}t}$, with the dashed line.
A comparison of the solid and dotted lines shows the effects of the
off resonance background contributions to reaction (\ref{eebnmbmn}),
and a comparison of the dotted and dashed lines shows the pure effect
of the off-mass-shell production of the $\ttbar$-pair.

How the off-resonance effects may distort the shape of differential cross
sections is shown in Figs~2, 3 and 4, 
where the angular distribution of a $\mu^+$ and the energy distributions 
of a $b$-quark and $\mu^+$ of reaction (\ref{eebnmbmn})
at $\sqrt{s}=500$ GeV are plotted, respectively.

\rput(7,-6){\scalebox{0.8 0.8}{\epsfbox{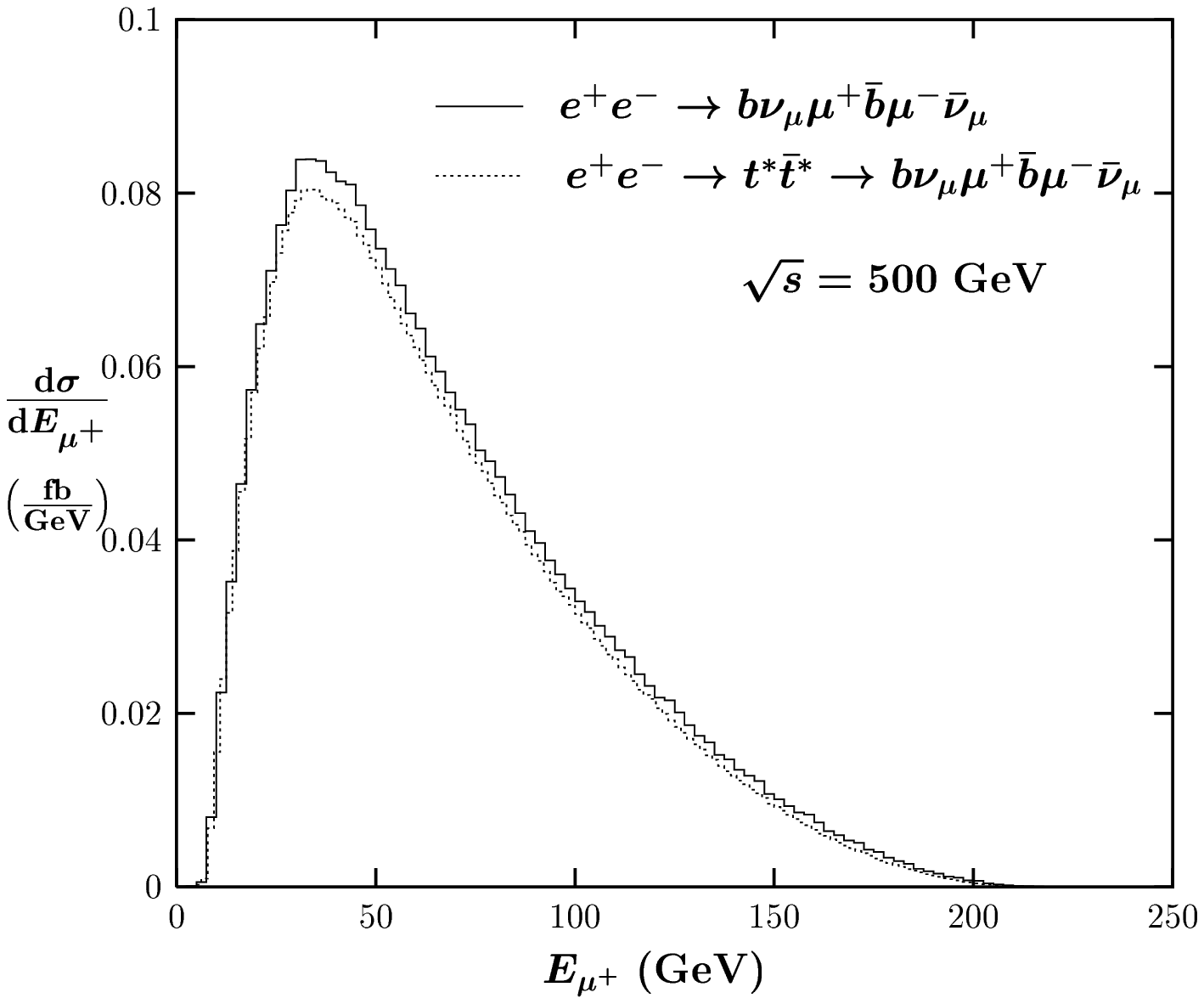}}}

\vspace*{9.0cm}

\bigskip
\bigskip
\begin{center}
{\small Figure~3. Energy distributions of a $\mu^+$ at $\sqrt{s}=500$ GeV.
\hspace*{2.cm}}
\end{center}

\section{SUMMARY AND OUTLOOK}

Top quark pair production and decay into 6 fermions in $\epm$  annihilation 
at c.m.s. energies typical for linear colliders has been studied with a 
program {\tt eett6f} and some sample results on reaction (\ref{eebnmbmn}) have
been presented. It has been shown that, although the two $\ttbar$ signal 
diagrams dominate total cross sections even at c.m.s. energies much above the 
$\ttbar$ threshold, the effects related to off-mass-shell production of the
$\ttbar$-pairs and the off resonance background may be relevant for the 
analyses of future precision data. 

\rput(7,-6){\scalebox{0.8 0.8}{\epsfbox{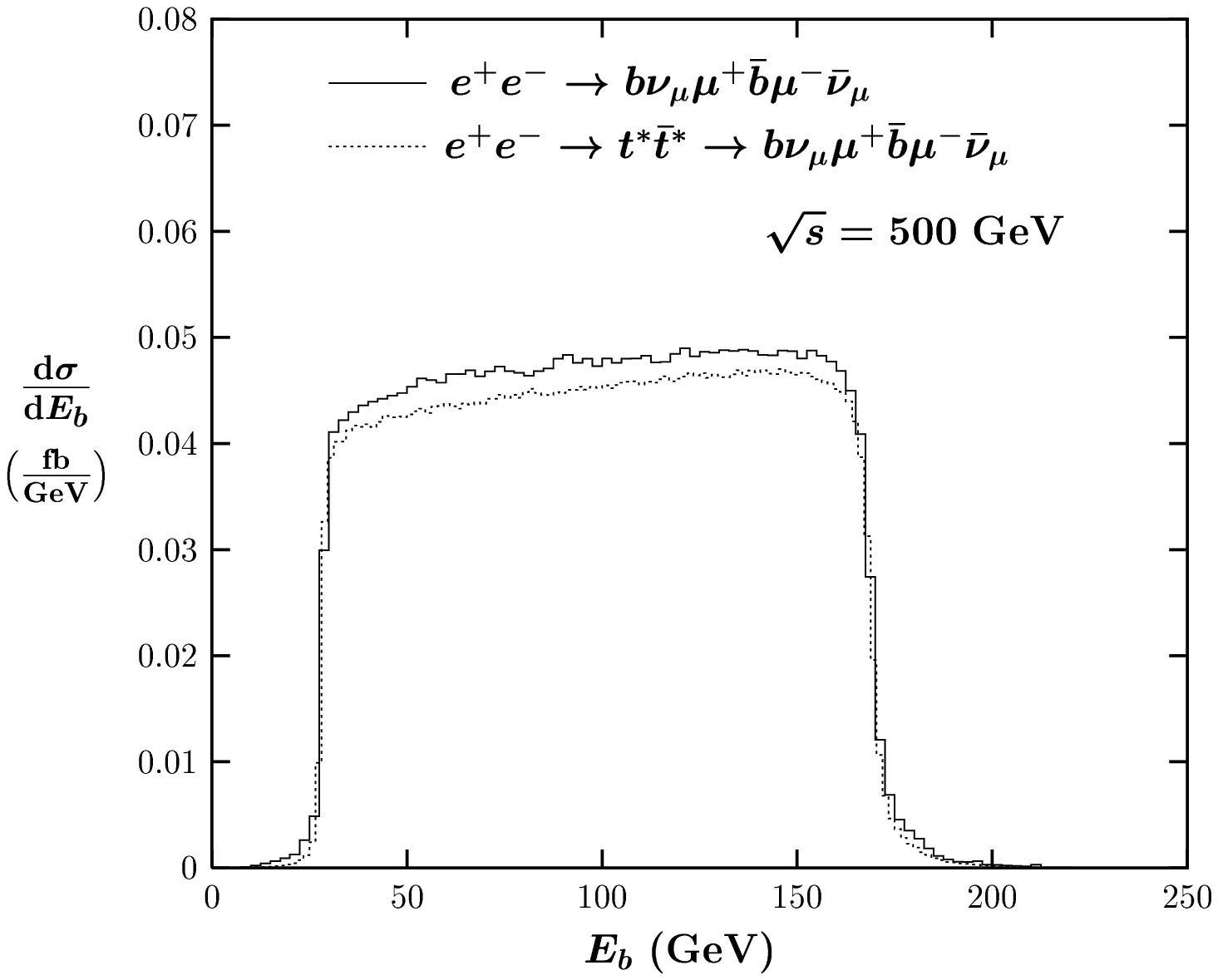}}}

\vspace*{9.0cm}

\bigskip
\bigskip
\begin{center}
{\small Figure~4. Energy distributions of a $b$-quark at $\sqrt{s}=500$ GeV.
\hspace*{2.cm}}
\end{center}

All reactions, except for those that include $e^{\pm}$
in the final state, have been implemented in {\tt eett6f v~1.0}. However,
the $t$-channel contributions to those reactions that are missing can be 
suppressed by even a small angular cut in the direction of initial beams, 
see {\it e.g.} \cite{BCK}. Some gluon contributions to $\eebudbdu$ are 
missing too, but their implementation into the program is underway.
Implementation of anomalous couplings is being done as well.

The fact that the total cross section of reactions (\ref{eesixf}) is dominated
by the doubly resonant signal justifies inclusion of leading radiative 
effects solely to the two signal diagrams. Work in this direction is planned 
in collaboration with the Zeuthen--Bielefeld group \cite{BZ}.

\end{document}